\documentclass[letter, scriptaddress, twocolumn, prd, showpacs,showkeys]{revtex4-1}

	\usepackage{amsmath}
        \usepackage{mathrsfs}
	\usepackage{amssymb}
	\usepackage{graphicx} 
	\usepackage{makeidx}
	\usepackage{amsfonts}
	\usepackage[ansinew]{inputenc}
	\usepackage[usenames, dvipsnames]{pstricks}
	\usepackage{epsfig}
	\usepackage{pst-grad} 
	\usepackage{pst-plot} 
	\usepackage[colorlinks, hyperindex]{hyperref}
	\hypersetup
	{
		colorlinks, %
		citecolor=blue, %
		linkcolor=Green, %
		urlcolor=blue, %
	}

\makeindex

\begin{document}

\title{\large{{\bf Unitarity in Reissner-Nordstr\"{o}m background: striding away from information loss}}}

\author{Arpit Das}
\email{ad13ms118@iiserkol.ac.in}
\author{Narayan Banerjee}
\email{narayan@iiserkol.ac.in}
\affiliation{Department of Physical Sciences, \\
Indian Institute of Science Education and Research Kolkata, \\
Mohanpur, West Bengal 741246, India.}

\begin{abstract}
We have shown analytically that radiation from a collapsing shell which leads to a charged black hole, whose exterior is described by the RN (Reissner-Nordstr\"om) metric (and hence the background spacetime is non-globally hyperbolic), is processed with a unitary evolution. For the analysis, we have used the Wheeler-deWitt formalism which in turn gave rise to a Schr\"{o}dinger-like wave equation. We showed the existence of unitarity by proving that the trace of the squared density matrix of the outgoing radiation, from a quantized massless scalar field, is unity and that the conservation of probability holds for the wave function of the system.   

\end{abstract}


\maketitle
\section{Introduction}
The information loss paradox, since its inception\cite{hawking1, hawking2}, has been open to many diverse interpretations\cite{wallace1, mathur1, polchinski1}. A traditional interpretation is: when a Schwarzschild black hole completely evaporates due to Hawking radiation the resultant spacetime becomes non-globally hyperbolic and hence quantum processes in such a spacetime would be non-unitary\cite{unruh1, unruh2}. This, as traditionally argued, would lead to information loss. On the contrary, we show that even in a non-globally hyperbolic spacetime, for instance in a Reissner-Nordstr\"{o}m (RN) background, unitarity can be achieved. We do this by adopting the Wheeler-deWitt formalism as in \cite{stojkovic1, vachaspati1}. In quantum mechanics, the evolution of pure states to mixed states may be understood as a non-unitary evolution. Recently, it has been shown by Saini and Stojkovic\cite{saini1}, in the context of a Schwarzchild black hole (and hence in a globally hyperbolic spacetime), that the evolution of quantum fields is actually unitary! They achieved this by showing that the traces of the density matrix and its square are unity for both the initial and the final states with proper normalization. However, they accomplished this by numerical estimates (over a period of finite proper time) which depend on the accuracy of the method and the reliability has to be ascertained carefully. \\

The present work shows for the first time, analytically, that the process of black hole radiation, even in a non-globally hyperbolic background, is unitary. This is proved using the consideration of density matrix, as in \cite{saini1}, but analytically. So the result is far more robust now. We further show, from quite an independent calculation, that the conservation of probability holds good in this process confirming an unitary evolution! Thus the result is now confirmed from two independent lines of approach. Furthermore, this is worked out for a non-globally hyperbolic background unlike the globally hyperbolic scenario as investigated in \cite{saini1}.\\

We work with an RN metric that includes an electric charge. The Schwarzchild case, considered in \cite{saini1}, is recovered as a special case by putting the charge $Q=0$. This generalization actually has important consequences. One can recover the extremal RN results by setting $|Q|=M$, where $M$ is the mass of the black hole. So, this work might have a profound significance in the context of string theory where an extremal RN black hole is a mainstay.

\section{The Model}
The model we will be studying comprises an infinitesimally thin collapsing charged spherical shell, with background metric $g_{\mu\nu}$ and a massless scalar field $\Phi$ whose dynamics we are interested in. The massless scalar field is assumed to couple to the gravitational field (originating from the presence of a non-trivial background metric), but not directly to the shell. We also have an asymptotic observer, sitting at the future null infinity, who is just there to register the outgoing flux with a detector and hence by assumption has very little or no interaction with ``shell-metric-scalar" system. Also, the observer is assumed not to significantly affect the evolution of the system and similarly for the system vis-a-vis the observer. The action for the whole system is then given by \cite{stojkovic1},
\begin{align}
S_{tot} = &\int d^4x\sqrt{-g}\left[-\frac{\mathcal{R}}{16\pi}+\frac{1}{2}(\partial_\mu\Phi)^2\right]-\sigma\int d^3\xi\sqrt{-\gamma} \nonumber\\ 
&+ S_{obs}, \label{Stot}
\end{align} 
where the first term is the Einstein-Hilbert term for the background metric $g_{\mu\nu}$, the second term is the action for the massless scalar field, the third term is shell's action in terms of its world-volume coordinates $\xi^a(a=0,1,2)$, the shell's tension $\sigma$ (or, shell's proper energy density per unit surface area) and the shell's induced world-volume metric $\gamma_{ab}$, which is given by,
\begin{align}
\gamma_{ab} &= g_{\mu\nu}\partial_a X^\mu\partial_b X^\nu, \label{gamma}
\end{align}
where $X^\mu(\xi^a)$ gives the location of the shell. The Roman indices run over the internal world-volume coordinates $\xi^a(a=0,1,2)$ while the Greek indices run over the usual spacetime coordinates.\\
\\
Lastly, $S_{obs}$ is the action for the observer.

\section{Spacetime Foliation-RN coordinates}
We consider that the mass and the charge is confined in an infinitesimally thin shell, so that for an exterior observer the distribution is spherical, whereas the inside of the shell is empty given by the Minkowski metric. The exterior of the shell is described by a Reissner-Nordst\"{o}m metric and its uniqueness is guaranteed by the charged version of Birkhoff's theorem\cite{birkhoff1, heusler1}. Thus, we have,
\begin{align}
ds^2_{out} = &-\left(1-\frac{2M}{r}+\frac{Q^2}{r^2}\right)dt^2 \nonumber \\ 
&+ \left(1-\frac{2M}{r}+\frac{Q^2}{r^2}\right)^{-1}dr^2 + r^2d\Omega_2^2,   \label{metric-out}\\
ds^2_{in} = &-dT^2 + dr^2 + r^2d\Omega_2^2,  \label{metric-in}\\
ds^2_{on-shell} = &-d\tau^2 + r^2d\Omega_2^2,  \label{met-onshell}
\end{align}
for $r>R(t)$, $r<R(t)$ and $r=R(t)$ respectively. $r$ is simply the radial coordinate and so $r=R(t)$ describes the shell. Furthermore, $R:=R(t)$,  $t$, $T$ and $\tau$ are the radius of the shell, time coordinate of the exterior observer, the time coordinate inside the shell and the proper time on the shell respectively. $d\Omega^2_2$ is the usual $S^2$ metric. \\
\\
One important thing to note here is that since RN coordinates leads to a coordinate singularity, at $R=R_H:=M+\sqrt{M^2-Q^2}$ (the event horizon), we might be in trouble using this for our analysis. However, note that for an asymptotic observer the event horizon is an infinitely red shifted surface. Hence, the observer can only notice the collapse of the shell approaching its event horizon in infinite time as per his time $t$. So, our analysis happens upto this limit which is relevant from an asymptotic viewpoint and RN coordinates are well behaved upto this limit, that is just outside the event horizon.
\\
\\
We consider timelike unit vectors $u^\alpha := \frac{dx_{out}^\alpha}{d\tau}$ and $v^\alpha := \frac{dx_{in}^\alpha}{d\tau}$, for $d s^2_{out}$ and $ds^2_{in}$ respectively. From their normalization, that is, $u^\alpha u_\alpha=-1$ and $v^\alpha v_\alpha=-1$, one obtains, at $r=R(t)$, $t_{\tau} = \frac{\sqrt{D+R_{\tau}^2}}{D}$, $T_{\tau} = \sqrt{1+R_{\tau}^2}$ and $T_t = \sqrt{D-(1-D)\frac{R_t^2}{D}}$. In the above expressions, a subscript indicates a differentiation w.r.t. that particular coordinate. $x^\alpha_{out}$ and $x^\alpha_{in}$ are the coordinates pertaining to $ds^2_{out}$ and $ds^2_{in}$ respectively. Also, $D:=1-\frac{2M}{R(t)}+\frac{Q^2}{R(t)^2}$.

\section{Mass of the shell}
Using Israel's formulation\cite{israel1}, the mass $M$ of the shell is (see also \cite{lopez1}),
\begin{align}
M &= 4\pi\sigma R^2\left[\sqrt{1+R_\tau^2}-2\pi\sigma R\right]+\frac{Q^2}{2R}, \label{M}
\end{align}
We show below that $M$ is a constant of motion. So, there is no conflict with the fact that $M$ is a constant of integration in the metric and is identified as the mass of the shell. Using the results given in \cite{wang1}, one can write,
\begin{align}
&\frac{R_{\tau\tau}}{\alpha} = \frac{Q^2}{8\pi\sigma R^4}+6\pi\sigma-\frac{2\alpha}{R}, \ \left(where, \ \alpha:=\sqrt{1+R_\tau^2}\right). \label{M_c}
\end{align}
Then, using $eq^ns$(\ref{M}) and (\ref{M_c}), 
\begin{align*}
M_\tau = & \ R_\tau\left[8\pi\sigma R(\alpha-2\pi\sigma R)-\frac{Q^2}{2R^2}\right] \\
&+R_\tau\left[4\pi\sigma R^2\left(\frac{Q^2}{8\pi\sigma R^4}-\frac{2\alpha}{R}+6\pi\sigma-2\pi\sigma\right)\right] =0.
\end{align*}
So, $M$ is a constant of motion.\\

Interpretations of $M$ can be looked at as follows. Suppose $R_\tau=0$ in $eq^n$(\ref{M}) (a static shell), then,
\begin{align}
M_{static} = 4\pi\sigma R^2-8\pi^2\sigma^2R^3 +\frac{Q^2}{2R}, \label{M-stat}
\end{align}
where the three terms represent the rest mass term, gravitational self-interaction term and the electrostatic self-interaction term respectively. If $R_\tau\neq 0$, then the term with $\sqrt{1+R_\tau^2}$ in $eq^n$ (\ref{M}) is the kinetic energy term. To get more intuition out of this, let us look at $M$ in the non-relativistic limit (where $R_\tau<<1$). By identifying constant mass $M_0:=4\pi\sigma R^2$, we have from $eq^n$(\ref{M}),
\begin{align}
M_{non-rel}=M_0+\frac{p^2}{2M_0} -\frac{M_0^2}{2R} +\frac{Q^2}{2R}, \label{M-nrel}
\end{align}
where $p:=\frac{1}{2}M_0R_\tau^2$ is the momentum of the particle with constant mass $M_0$ moving in a gravitational and electrostatic potential. Here the second term represents the kinetic energy. One can clearly identify $eq^n$(\ref{M-nrel}) as the Hamiltonian of a non-relativistic particle moving under the influence of a gravitational and electrostatic potential (see \cite{wang1}). From relativistic perspectives, note that (see \cite{vachaspati1}),
\begin{align}
M_{rel} = \frac{M_0}{\sqrt{1-R_T^2}} -\frac{M_0^2}{2R} + \frac{Q^2}{2R}, \label{M-rel}
\end{align}  
which is the Hamiltonian of a relativistic particle with rest mass $M_0$ moving in a gravitational and electrostatic potential.\\
\\
Since, we have shown that $M$ is a constant of motion, by the above interpretation of $M$, we have the following identification,
\begin{align}
\mathcal{H}_{shell}\equiv M, \label{H-s}
\end{align}
where $H_{shell}$ is the Hamiltonian of the shell to be treated classically.

\section{Action for the shell}
The form of the action, for the shell, is taken as
\begin{align}
S_{shell} = -\int dT \ \left[4\pi\sigma R^2\left[\sqrt{1-R_T^2}-2\pi\sigma R\right] + \frac{Q^2}{2R}\right]. \label{S-shell}
\end{align}

The corresponding Lagrangian yields the conjugate momentum, for the shell, as,
\begin{align}
\Pi_{shell} = \frac{\partial\mathcal{L}_{shell}}{\partial R_T} = 4\pi\sigma R^2\left(\frac{R_T}{\sqrt{1-R_T^2}}\right). \label{P-shell}
\end{align}

Then, the Hamiltonian is,
\begin{align}
\mathcal{H}_{shell} = & \ \Pi_{shell}R_T - \mathcal{L}_{shell} \nonumber\\
= & \ 4\pi\sigma R^2\left[\sqrt{1+R_\tau^2}-2\pi\sigma R\right]+\frac{Q^2}{2R}.
\label{H-shell}
\end{align}
This matches with $M$ as expressed in $eq^n(\ref{M})$. So, the action in $eq^n$(\ref{S-shell}) is consistent (since, this action gives the correct $\mathcal{H}_{shell}$ as expressed in $eq^n$(\ref{H-s})). In terms of time $t$, (using the expression for $T_t$) $S_{shell}$ becomes,
\begin{align}
S_{shell} = &-\int dt \ \left[4\pi\sigma R^2\left[\sqrt{D-\frac{R_t^2}{D}}\right]\right] \nonumber\\
&+\int dt \ \left[4\pi\sigma R^2\left[2\pi\sigma R\sqrt{D-\frac{1-D}{D}R_t^2}\right]\right] \nonumber\\
&-\int dt \ \left[\frac{Q^2}{2R}\sqrt{D-\frac{1-D}{D}R_t^2}\right]. \label{S-st}
\end{align}
The Conjugate Momentum and Hamiltonian (in terms of $t$) are,
\begin{align}
\Pi_{shell} = & \ \frac{\partial\mathcal{L}_{shell}}{\partial R_t} \nonumber\\ 
= & \ \frac{4\pi\sigma R^2R_t}{\sqrt{D}}\left[\frac{1}{\sqrt{D^2-R_t^2}}-\frac{2\pi\sigma R(1-D)}{\sqrt{D^2-(1-D)R_t^2}}\right] \nonumber\\
&+ \frac{4\pi\sigma R^2R_t}{\sqrt{D}}\left[\frac{Q^2(1-D)}{8\pi\sigma R^3\sqrt{D^2-(1-D)R_t^2}}\right], \label{P-st}
\end{align}

\begin{align}
\mathcal{H}_{shell} = & \ \Pi_{shell}R_t - \mathcal{L}_{shell} \nonumber\\ 
= & \ 4\pi\sigma D^{3/2}R^2\left[\frac{1}{\sqrt{D^2-R_t^2}}-\frac{2\pi\sigma R}{\sqrt{D^2-(1-D)R_t^2}}\right] \nonumber\\
&+ 4\pi\sigma D^{3/2}R^2\left[\frac{Q^2}{8\pi\sigma R^3\sqrt{D^2-(1-D)R_t^2}}\right]. \label{H-st}
\end{align}

\section{Incipient Limit}
The incipient limit, $R\rightarrow R_H$, is the limit when the radius of the shell approaches the event horizon,
\begin{align}
R_H &= M+\sqrt{M^2-Q^2}. \label{R-h1}
\end{align}
From $eq^n$(\ref{P-st}) and $eq^n$(\ref{H-st}) we note that, in the incipient limit, 
\begin{align}
&\Pi_{shell} = \frac{4\pi\mu R^2 R_t}{\sqrt{D}\sqrt{D^2-R_t^2}}, \label{P-sti}
\end{align}
\begin{align}
&\mathcal{H}_{shell} = \frac{4\pi D^{3/2}\mu R^2}{\sqrt{D^2-R_t^2}}, \label{H-sti}
\end{align}
where, $\mu:=\sigma\left(1-2\pi\sigma R_H+\frac{Q^2}{8\pi\sigma R_H^3}\right)$. So we have,
\begin{align}
\mathcal{H}_{shell} = [(D\Pi_{shell})^2+D(4\pi\mu R^2)^2]^{1/2} \equiv [q^2+m^2]^{1/2}, \label{H-strel}
\end{align}
where $q^2:=(D\Pi_{shell})^2$ and $m^2:=D(4\pi\mu R^2)^2$.\\
\\
$Eq^n$(\ref{H-strel}) shows that $\mathcal{H}_{shell}$ is the Hamiltonian of a relativistic particle with a position dependent mass. So, that is how the shell behaves in the incipient limit. Let us now show that in this limit also, $\mathcal{H}_{shell}$ is a constant of motion. Since, $\frac{d\mathcal{H}_{shell}}{d\tau}=\frac{\partial \mathcal{H}_{shell}}{\partial\tau}$, we have,
\begin{align}
&\frac{d}{d\tau}\left(4\pi\mu\frac{D^{3/2}R^2}{\sqrt{D^2-R_t^2}}\right) = 0 \nonumber\\
leading \ to, \ & \frac{D^{3/2}R^2}{\sqrt{D^2-R_t^2}} = \frac{\mathcal{H}_{shell}}{4\pi\mu} =: h \ (a \ constant), \label{h} \\
&(as \ \tau \ doesn't \ appear \ explicitly \ in \ \mathcal{H}_{shell}). \nonumber
\end{align}

We can arrive at these expressions independently using an alternative approach (see appendix)

Classically, we have from $eq^n$(\ref{h}) and $T_t$,
\begin{align}
&R_t = \pm D\sqrt{1-\frac{D R^4}{h^2}} \approx \pm D\left(1-\frac{1}{2}\frac{DR^4}{h^2}\right) \approx \pm D \label{R-ti} \\
&(as \ R\rightarrow R_H), \nonumber\\
&T_t = D\sqrt{1+(1-D)\frac{R^4}{h^2}}, \label{T-ti}
\end{align}
where solving $eq^n(\ref{R-ti})$ in terms of $t$ will give us the classical behaviour of the shell as the event horizon is approached.

\section{Non-extremal case}
The horizons (outer and inner) of the charged shell are given by,
\begin{align*}
R_{\pm}&= M {\pm} \sqrt{M^2-Q^2}.
\end{align*}

$D$ can be written as,
\begin{align}
D = \left(1-\frac{R_+}{R}\right)\left(1-\frac{R_-}{R}\right) =: D_+D_-. \label{D-i1}
\end{align}
In the incipient limit, $D\rightarrow 0$ and hence, $D_+\rightarrow 0$ (as $R(t)$ approaches the event horizon $R_H=R_+$). Thus, for $D_+\rightarrow 0$, we have $D_-\rightarrow 1-\frac{R_-}{R_+}=:\left.D_-\right|_i$. Then, in this limit, $R_t\approx\pm D=\pm D_+D_-$. Solving for $R(t)$ we get (from $eq^n$(\ref{R-ti}) and $eq^n$(\ref{D-i1})),
\begin{align}
\pm 1=\frac{1}{D_-}\frac{R}{R-R_+}\frac{dR}{dt} &\approx \frac{1}{\left.D_-\right|_i}\frac{R_+}{R-R_+}\frac{dR}{dt} \nonumber\\
&(upto \ leading \ order) \nonumber\\
integrating \ w.r.t. \ t, \ & R_+ ln\left(\frac{R_f-R_+}{R_0-R_+}\right) = \pm t_f \left.D_-\right|_i \nonumber\\ 
&(R_0:=R(0) \ and \ R_f:=R(t_f)) \nonumber\\
and \ thus, \ & R_f = R_+ + (R_0-R_+) \ e^{\pm \left.D_-\right|_i t_f/R_+}, \label{Rf-ti1}
\end{align}
where the lower limit of integration w.r.t. $t$ is $t=0$ and the upper limit is $t=t_f$.\\
\\
As $R_f\rightarrow R_+$ and $t_f>0$ along with $\left.D_-\right|_i>0$, we see that, $t_f\rightarrow\infty$. So, the negative sign for $R(t)$ describes a collapsing model in the incipient limit. $Eq^n$(\ref{Rf-ti1}) also shows that for an asymptotic observer, the formation of the event horizon takes infinite time implying that the event horizon is an infinite red shifted surface, which matches with the classical result, as stated earlier while choosing the RN coordinates. 

\section{Action for the scalar field $\Phi$}
The action for the scalar field $\Phi$ is written as a sum of the actions,
\begin{align}
S_{\Phi} =& \left.S_{\Phi}\right)_{in} + \left.S_{\Phi}\right)_{out} \nonumber \\
= & \ 2\pi\int dt\left[-(\partial_t\Phi)^2\left(\int_0^R dr \ r^2\frac{1}{T_t}\right)\right] \nonumber\\
&+2\pi\int dt\left[(\partial_r\Phi)^2\left(\int_0^R dr \ r^2 \ T_t\right)\right] \nonumber\\
&+2\pi\int dt\left[-(\partial_t\Phi)^2\left(\int_R^\infty dr \ r^2\frac{1}{1-\frac{2M}{r}+\frac{Q^2}{r^2}}\right)\right] \nonumber\\
&+2\pi\int dt\left[(\partial_r\Phi)^2\left(\int_R^\infty dr \ r^2\left(1-\frac{2M}{r}+\frac{Q^2}{r^2}\right)\right)\right], \label{S-phicom}
\end{align}
where the limits of the integration w.r.t. $r$ for $S_{\Phi})_{in}$ are from $0$ to $R$ while for $S_{\Phi})_{out}$ are from $R$ to $\infty$.
\\
\\
$T_t\rightarrow D \ (upto \ leading \ order)$ as $R\rightarrow R_H$ (from $eq^n$(\ref{T-ti})). So,
\begin{align*}
\lim_{R\rightarrow R_H}\frac{T_t}{1-\frac{2M}{r}+\frac{Q^2}{r^2}}=\frac{R^2-2MR+Q^2}{r^2-2Mr+Q^2}\frac{r^2}{R^2} = 0.
\end{align*}
$T_t$ vanishes faster than $\left(1-\frac{2M}{r}+\frac{Q^2}{r^2}\right)$ in the limit $R\rightarrow R_H$. Thus, for coefficients of $-(\partial_t\Phi)^2$, the $\frac{1}{T_t}$ term dominates and for coefficients of $(\partial_r\Phi)^2$, the term which dominates is $\left(1-\frac{2M}{r}+\frac{Q^2}{r^2}\right)$. Thus, in the incipient limit,
\begin{align}
S_{\Phi} \rightarrow & \ 2\pi\int dt\left[-\frac{1}{D}\int_0^{R_H}dr \ r^2 (\partial_t\Phi)^2\right] \nonumber\\
&+2\pi\int dt\left[\int_{R_H}^{\infty}dr \ r^2 \left(1-\frac{2M}{r}+\frac{Q^2}{r^2}\right)(\partial_r\Phi)^2\right]. \label{S-phii}
\end{align}

\section{Mode expansion for $\Phi$}
For a massless scalar field $\Phi$, one can easily check from its equation of motion, that is $\partial^2\Phi = 0$, that for $r<R(t)$ (from $\left.S_{\Phi}\right)_{in}$),
\begin{align}
\frac{\partial^2\Phi}{\partial r^2} + \frac{2}{r}\frac{\partial\Phi}{\partial r} = \frac{1}{T_t^2}\frac{\partial^2\Phi}{\partial t^2}-\frac{T_{tt}}{T_t^3}\frac{\partial\Phi}{\partial t}, \label{wavein}
\end{align}
where, $T_t$, and hence its powers and derivatives w.r.t. $t$, are independent of $r$.\\
\\ 
Similarly, for $r>R(t)$, we have (from $\left.S_{\Phi}\right)_{out}$)),
\begin{align}
&\left(1-\frac{2M}{r}+\frac{Q^2}{r^2}\right)^2\frac{\partial^2\Phi}{\partial r^2} \nonumber\\
&+ \frac{2(r-M)}{r^2}\left(1-\frac{2M}{r}+\frac{Q^2}{r^2}\right)\frac{\partial\Phi}{\partial r} = \frac{\partial^2\Phi}{\partial t^2}. \label{waveout}
\end{align}
From $eq^n$(\ref{wavein}) and $eq^n$(\ref{waveout}), we have the following mode expansion (due to the separability property of the above equations),
\begin{align}
\Phi(r,t) = \sum_k a_k(t)f_k(r), \label{phi-rt}
\end{align}
where $a_k(t)$ are the modes and $f_k(r)$ are real-valued smooth functions of r.
\\
\\
$S_\Phi$ in terms of modes $a_k$ is (as $R\rightarrow R_H$),
\begin{align}
S_\Phi = \int dt \ \sum_{k,k^{'}} \left[-\frac{1}{2D}\frac{da_k}{dt}A_{kk^{'}}\frac{da_{k^{'}}}{dt}+\frac{1}{2}a_kB_{kk^{'}}a_{k^{'}}\right], \label{S-phim}
\end{align}
with the following definitions for $A_{kk^{'}}$ and $B_{kk^{'}}$,
\begin{align}
&A_{kk^{'}} := 4\pi\int_0^{R_H}dr \ r^2f_k(r)f_{k^{'}}(r), \label{Akk} \\
&B_{kk^{'}} := 4\pi\int_{R_H}^{\infty}dr \ r^2\left(1-\frac{2M}{r}+\frac{Q^2}{r^2}\right)f^{'}_k(r)f^{'}_{k^{'}}(r), \label{Bkk}
\end{align}
where, $f_k^{'}(r) := \frac{\partial f_k(r)}{\partial r}$. Note that, both $A_{kk^{'}}$ and $B_{kk^{'}}$ are independent of $r$ and $t$ (as no $R(t)$ appears in them). \\
\\
The cojugate momenta, $\pi_k$s (to the modes $a_k$), are defined as,
\begin{align}
\pi_k := \frac{\partial \mathcal{L}_{\Phi}}{\partial \dot{a_k}} \equiv -i\frac{\partial}{\partial a_k}, \label{pk}  
\end{align}
where, $\dot{a}_k := \frac{da_k}{dt}$, and from $eq^n$(\ref{S-phim}), we have (with $\mathcal{L}_{\Phi}$ defined as the Langrangian for $\Phi$),
\begin{align}
\mathcal{L}_{\Phi} &= \sum_{k,k^{'}} \left[-\frac{1}{2D}\dot{a}_k A_{kk^{'}}\dot{a}_{k^{'}}{dt}+\frac{1}{2}a_kB_{kk^{'}}a_{k^{'}}\right], \label{L-phim1} \\
\mathcal{L}_{\Phi} &= -\frac{1}{2D}(\mathbf{\dot{a}}^T \mathbf{A}\mathbf{\dot{a}}) + \frac{1}{2}(\mathbf{a}^T \mathbf{B}\mathbf{a}), \label{L-phim2} 
\end{align}
where, $\mathbf{A}$ and $\mathbf{B}$ are non-singular linear operators, such that, $A_{kk^{'}}\in\mathbf{A}$ and $B_{kk^{'}}\in\mathbf{B}$ in the chosen bases, say $\lbrace \dot{a_k}\rbrace$ and $\lbrace a_k\rbrace$ respectively. In the basis $\lbrace a_k\rbrace$, $\mathbf{a}$ is a column vector, such that, $a_k\in\mathbf{a}$. One can express $\mathbf{\dot{a}}$ in a similar way in the basis $\lbrace \dot{a_k}\rbrace$.\\
\\  
For the Hamiltonian of $\Phi$, $\mathcal{H}_{\Phi}$, we get,
\begin{align}
\mathcal{H}_{\Phi} = &\sum_{k} \pi_k\dot{a}_k - \mathcal{L}_{\Phi}\nonumber\\ 
= &\sum_{k,k^{'}} \left[\frac{1}{2D}\dot{a}_k A_{kk^{'}}\dot{a}_{k^{'}}{dt}+\frac{1}{2}a_kB_{kk^{'}}a_{k^{'}}\right] \label{H-phim1} \\
= & \ \frac{D}{2}(\mathbf{\Pi}^T\mathbf{A}^{-1}\mathbf{\Pi})+\frac{1}{2}(\mathbf{a}^T \mathbf{B}\mathbf{a}), \label{H-phim2}
\end{align}
where $\mathbf{\Pi}$ is a column vector, such that, $\pi_k\in\mathbf{\Pi}$, in a chosen basis say $\lbrace \pi_k\rbrace$ and $\mathbf{A}^{-1}$ denotes the inverse of $\mathbf{A}$.\\
\\ 
$\mathbf{B}$ and $\mathbf{A}$ are real and symmetric infinite dimensional matrices and hence are self-adjoint. Thus, by the {\itshape{Spectral Theorem}}, there exists orthonormal bases of position space and momentum space consisting of respective eigenvectors of $\mathbf{B}$ and $\mathbf{A}$. Furthermore, all the eigenvalues are real. Say the bases for position space and momentum space are $\lbrace b_k\rbrace$ and $\lbrace \dot{b}_k\rbrace$ respectively (where, each $b_k$ is a linear combination of the original basis vectors $a_k$ and each $\dot{b}_k$ is a linear combination of the original basis vectors $\dot{a}_k$).

\section{The Schr\"{o}dinger-like wave equation}
If we analyze the equation for one eigenvector $b\in\lbrace b_k\rbrace$, then our conclusion will be the same for all other eigenvectors (see \cite{stojkovic1}). So, we will solve the Schr\"{o}dinger-like wave equation for a wave functional $\Psi(\lbrace b_k\rbrace,t)$ (see appendix), which by the above assumption of equivalence is now a wave function $\psi(b,t)$. Hence, $\psi(b,t)\equiv\Psi(\lbrace b_k\rbrace,t)$. Thus, using $eq^n$(\ref{L-phim2}), we write the Schr\"{o}dinger-like wave equation (for a single eigenvector $b$) as,
\begin{align}
\left[-\left(1-\frac{2M}{R}+\frac{Q^2}{R^2}\right)\frac{1}{2\alpha}\frac{\partial^2}{\partial b^2}+\frac{1}{2}\beta b^2\right]\psi(b,t) = i\frac{\partial \psi(b,t)}{\partial t}, \label{Schr-t}
\end{align}
where, $\alpha$ and $\beta$ are the eigenvalues of $\mathbf{A}$ and $\mathbf{B}$ respectively.\\ 
\\
Let us define a new time parameter,
\begin{align}
&\eta := \int_0^t dt \ \left(1-\frac{2M}{R}+\frac{Q^2}{R^2}\right) \label{eta1} \\
leading \ to, \ &\frac{\partial \eta}{\partial t} = D, \label{eta2}
\end{align}
and write $eq^n$(\ref{Schr-t}) as
\begin{align}
\left[-\frac{1}{2\alpha}\frac{\partial^2}{\partial b^2}+\frac{\beta}{2D}b^2\right]\psi(b,\eta) = i\frac{\partial \psi(b,\eta)}{\partial \eta}. \label{Schr-eta}
\end{align}
Define,
\begin{align}
\omega^2(\eta) := \left(\frac{\beta}{\alpha}\right)\frac{1}{D} =:\frac{\omega_0^2}{D}. \label{omega} 
\end{align}
Then, $eq^n$(\ref{Schr-eta}) becomes,
\begin{align}
\left[-\frac{1}{2\alpha}\frac{\partial^2}{\partial b^2}+\frac{1}{2}\alpha\omega^2(\eta)b^2\right]\psi(b,\eta) = i\frac{\partial \psi(b,\eta)}{\partial \eta}, \label{Schr-sho}
\end{align}
where, we have chosen to set $\eta(t=0)=0$. Observe that, $eq^n(\ref{Schr-sho})$ is a time dependent Simple Harmonic Oscillator (SHO) equation with $\omega(\eta)$ as the SHO's frequency. \\

In the incipient limit (using  $eq^n(\ref{D-i1})$ and  $eq^n(\ref{R-ti})$),  
\begin{align}
\frac{dD}{dt} &= \left[D_-\frac{R_+}{R^2}+D_+\frac{R_-}{R^2}\right] \approx -\left.D_-\right|_i\frac{R_+}{R_+^2}D = -\frac{\left.D_-\right|_i D}{R_+}. \ \ \  \label{Dt-i}
\end{align}
Integrating $eq^n(\ref{Dt-i})$ w.r.t. $t$ one gets (as $R\rightarrow R_H$),
\begin{align}
D &=1-\frac{2M}{R(t)}+\frac{Q^2}{R(t)^2} \sim e^{-\left.D_-\right|_it/R_+} \label{D-i3}.
\end{align}
From $eq^n(\ref{D-i3})$ we see that at late time, $1-\frac{2M}{R(t)}+\frac{Q^2}{R(t)^2} \sim e^{-\left.D_-\right|_it/R_+}$. Since we are interested in the incipient limit, that is, in late times of the collapsing process, we can choose the behaviour of $R(t)$ at early times as per our convenience for simplifying calculations. So we choose both past and future behaviour of $R(t)$ to be stationary. We can take the metric to be flat for all $t\in(-\infty,0)$. Stationarity in future can be achieved by taking a cut-off time $t_f$ for the collapse and then allowing $t_f\rightarrow\infty$, thus going into the continual collapse case till black hole formation. Thus,
\begin{align}
D&= 
\begin{cases}
1, \ \ \ \ \ \ \ \ \ \ \ \ \ \ \ \ \ for \ \ \ t\in(-\infty,0) & \\
e^{-\left.D_-\right|_it/R_+}, \ \ \ \ for \ \ \ t\in(0,t_f) & \\
e^{-\left.D_-\right|_it_f/R_+}, \ \ \ for \ \ \ t\in(t_f,\infty). &
\end{cases} \label{D-choice1}
\end{align} 
The above choice of $R(t)$ may seem problematic as $\frac{dR}{dt}$ is discontinuous at $0$ and $t_f$, but references  \cite{stojkovic1, greenwood1} show that the particle production by the collapsing shell happens in the range, $0<t<t_f$ and in the $t_f\rightarrow\infty$ regime, all the solutions obtained are well-behaved. So with the above considerations, the wavefunction $\psi$ would capture the whole collapse scenario, and in the limit of $t_f\rightarrow\infty$ or $R(t)\rightarrow R_H$, black hole formation occurs.\\

We note that, at early times, $t\in(-\infty,0)$, the spacetime is Minkowski and hence the initial vacuum states at $\mathcal{J^{-}}$ (past null infinity) are (\footnote{The intuition behind identifying the states at $\mathcal{J^{-}}$ and states with $t\in(-\infty,0)$ comes from the fact that the observer is at $r\rightarrow\infty$ and at an early time at $t\rightarrow -\infty$ which is $\mathcal{J^{-}}$ and at late times he is at $t\rightarrow\infty$ which is $\mathcal{J^{+}}$.}) just the simple harmonic oscillator ground states (this can be seen from the form of $eq^n(\ref{Schr-sho})$, which with $\eta=0$, is the SHO equation). Thus,
\begin{align}
\psi_0(b) := \psi(b,\eta=0) = \left(\frac{\alpha\omega_0}{\pi}\right)^{1/4}e^{-m\omega_0b^2/2}, \label{phi-gs}
\end{align}
where $\psi_0(b)$ represents the SHO ground state and $\lbrace\psi_n(b)\rbrace$ will represent the SHO basis states at early times.\\
\\
$Eq^n(\ref{phi-gs})$ suggests that $\omega_0$ defined in $eq^n(\ref{omega})$ is the ground state frequency associated with the initial vacuum state.\\
\\
With the aid of $eq^n(\ref{phi-gs})$, the exact solution to $eq^n(\ref{Schr-sho})$ is, 
\begin{align}
\psi(b,\eta) = e^{i\chi(\eta)}\left[\frac{\alpha}{\pi\zeta^2}\right]^{1/4}\exp\left[i\left(\frac{\zeta_\eta}{\zeta}+\frac{i}{\zeta^2}\right)\frac{\alpha b^2}{2}\right], \label{psi-soln}
\end{align}
where $\zeta$ is the solution of the equation,
\begin{align}
\zeta_{\eta\eta} + \omega^2(\eta)\zeta = \frac{1}{\zeta^3}, \label{dif1}
\end{align}
with the following initial conditions,
\begin{align}
&\zeta(0)=\frac{1}{\sqrt{\omega_0}}, \label{dif2} \\
&\zeta_\eta(0) = 0 \label{dif3},
\end{align}
and, $\chi(\eta)$ is given by,
\begin{align}
\chi(\eta) := -\frac{1}{2}\int^{\eta}_0\frac{d\eta^{'}}{\zeta^2(\eta^{'})}. \label{dif4}
\end{align}
Equations of the form $eq^n(\ref{Schr-sho})$ have been extensively studied in \cite{dantas1, lewis1, lewis2, pedrosa1, kolopanis1}.\\ \\

From $eq^n(\ref{omega})$, $eq^n(\ref{D-i3})$ and $eq^n(\ref{D-choice1})$, we have (for $t>0$),
\begin{align}
\omega(\eta(t)) = e^{\left.D_-\right|_it/2R_+}\omega_0 .
\label{omega1}
\end{align}
Using \ $eq^n(\ref{eta2})$ and $eq^n(\ref{omega1})$, 
\begin{align}
\Omega(t) &= \left(\left.\frac{\partial\eta}{\partial t}\right|_{t>0}\right)\omega(\eta) = e^{-\left.D_-\right|_it/2R_+}\omega_0 ,
\label{omega2}
\end{align}
where $\Omega(t)$ is defined to be the frequency w.r.t. time $t$.\\

We note that at early times ($\mathcal{J^{-}}$), the states are the initial vacuum states of SHO described by $\psi_0(b)$. With time, the frequency of the states $\Omega(t)$ evolve (as per $eq^n(\ref{omega2})$) and more states get excited. Finally, when the observer measures them at $\mathcal{J^{+}}$ (future null infinity), that is for some $t\in(t_f,\infty)$, we have (following the evolution as per the Schr\"{o}dinger picture\cite{sakurai1}),
\begin{align}
\psi(b,t) = \sum_n c_n(t)\phi_n(b), \label{psigen}
\end{align}
where $c_n(t)$ are the probability amplitudes and the final SHO states $\lbrace\phi_n(b)\rbrace$ are with the frequency $\Omega_f=\Omega(t_f)$ (a constant), given by,
\begin{align}
\phi_n(b) = \left(\frac{\alpha\Omega_f}{\pi}\right)^{1/4}\frac{e^{-\alpha\Omega_f b^2/2}}{\sqrt{2^n n!}}H_n(\sqrt{\alpha\Omega_f}b). \label{phin123}
\end{align}
Here, $H_n$ are the Hermite polynomials. Note that,
\begin{align}
\Omega(t_f) = e^{-\left.D_-\right|_it_f/2R_+}\omega_0;
\label{omega3}
\end{align}
$c_n$ can be computed from an overlap integral as (see appendix),
\begin{align}
c_n = \begin{cases} 
\frac{(-1)^{n/2}e^{i\chi}}{(\Omega_f\zeta^2)^{1/4}}\sqrt{\frac{2}{P}}\left(1-\frac{2}{P}\right)^{n/2}\frac{(n-1)!!}{\sqrt{n!}}, \ \ \ \ for \ \ \ even \ n & \\   
0, \ \ \ \ \ \ \ \ \ \ \ \ \ \ \ \ \ \ \ \ \ \ \ \ \ \ \ \ \ \ \ \ \ \ \ \ \ \ \ \ \ \ \ \ for \ \ \ odd \ n, &
\end{cases} \label{cn}
\end{align}
where $P:=1-\frac{i}{\Omega_f}\left(\frac{\zeta_\eta}{\zeta}+\frac{i}{\zeta^2}\right)$.

\section{Unitarity from Density Matrix}
We compute the density matrices, $\hat{\rho}_{i}$ and $\hat{\rho}_{f}$, for the initial ($\mathcal{J^{-}}$) and the final ($\mathcal{J^{+}}$) states respectively. We can write the $\hat{\rho}_{i}$ and $\hat{\rho}_{f}$ as (see \cite{saini1, saini2}),
\begin{align}
\hat{\rho}_i &= \sum_{m,n}l_m l_n^{*}|\psi_m\rangle\langle\psi_n|, \label{rhoi} \\
\hat{\rho}_f &= \sum_{m,n}c_m c_n^{*}|\phi_m\rangle\langle\phi_n|, \label{rhof}
\end{align} 
where, $l_n$ and $c_n$ are the probability amplitudes appearing in the intial and final states respectively. \\

Since initially the system was in the SHO eigenstates $\lbrace\psi_n\rbrace$ and the wavefunction is normalized, we have,
\begin{align}
Tr(\hat{\rho}_i) = 1. \label{trrhoi}
\end{align}
From $eq^n(\ref{cn})$, with $\kappa := \left|1-\frac{2}{P}\right|$, one has
\begin{align}
Tr(\hat{\rho}_f) &= \sum_{even \ n} |c_n|^2 \nonumber\\
&= \frac{2}{\sqrt{\Omega_f\zeta^2}|P|}\sum_{even \ n}\frac{(n-1)!!}{n!}\kappa^n \nonumber\\
&= \frac{2}{\sqrt{\Omega_f\zeta^2}|P|}\frac{1}{\sqrt{1-\kappa^2}} \nonumber\\
&= \frac{2}{\sqrt{\Omega_f\zeta^2}|P|}\frac{1}{\sqrt{1-\left|1-\frac{2}{P}\right|^2}}. \label{trrhof1}
\end{align}
$P$ has been computed explicitly and used in $eq^n(\ref{trrhof1})$ to obtain (see appendix),
\begin{align}
Tr(\hat{\rho}_f) = 1. \label{trrhof2}
\end{align}
By $eq^n(\ref{trrhof2})$, we have shown that the necessary condition for the unitary evolution of states holds. For the sufficiency, we compute $Tr(\hat{\rho}_f^2)$. From $eq^n(\ref{rhof})$, 
\begin{align}
\hat{\rho}_f &= \sum_{m,n}c_m c_n^*|\phi_m\rangle\langle\phi_n| \nonumber\\
leading \ to, \ \hat{\rho}_f^2 &= \left(\sum_{m,n}c_m c_n^*|\phi_m\rangle\langle\phi_n|\right)\left(\sum_{i,j}c_i c_j^*|\phi_i\rangle\langle\phi_j|\right) \nonumber\\
&= \sum_{m,n,i,j} c_m c_i c_n^* c_j^*|\phi_m\rangle\langle\phi_n|\phi_i\rangle\langle\phi_j| \nonumber\\
&= \sum_{m,n,j} c_m c_j^*|c_n|^2|\phi_m\rangle\langle\phi_j| \nonumber\\
&= \sum_{m,j} c_m c_j^*|\phi_m\rangle\langle\phi_j|\left(\sum_n|c_n|^2\right) \nonumber\\
&= \sum_{m,j} c_m c_j^*|\phi_m\rangle\langle\phi_j| \nonumber\\
&\left(as, \left(\sum_n|c_n|^2\right)=1 \ by \ eq^n(\ref{trrhof2})\right) \nonumber\\
&=\hat{\rho}_f. \label{trrho2f}
\end{align}
Thus, by $eq^n(\ref{trrho2f})$ we get,
\begin{align}
Tr(\hat{\rho}_f^2) = Tr(\hat{\rho}_f) = 1. \label{idemp}
\end{align}
Analytically, the idempotency of the final density matrix holds indicating a pure state to pure state transition.

\section{Unitarity from Conservation of Probability}
The probability current 4-vector $J^\mu$  is defined as,
\begin{align}
&J^0 = |\psi|^2, \label{j1} \\
&\vec{J} = \frac{1}{2\alpha i}[\psi^{*}\vec{\nabla}\psi - \psi\vec{\nabla}\psi^{*}]. \label{j2}
\end{align}
As $b$ is an eigenfunction of $\mathbf{B}$ which is independent of spatial coordinate $x^i$, we conclude that $\vec{J}=\vec{0}$. This suggests,
\begin{align}
\nabla_\mu J^\mu = \frac{\partial |\psi|^2}{\partial t_{obs}}. \label{j3}
\end{align} 
Writing $t_{obs} = t$ (for the observer's time coordinate), we have (from equation (\ref{eta2})),
\begin{align}
&\nabla_\mu J^\mu =  \frac{\partial |\psi|^2}{\partial t} = \frac{\partial |\psi|^2}{\partial\eta}\frac{\partial\eta}{\partial t} = D\frac{\partial |\psi|^2}{\partial\eta} \nonumber \\
&For, \ R\rightarrow R_H, \ \nabla_\mu J^\mu = 0 \ \ \ (as, \ D\rightarrow 0)
\label{j4}
\end{align}
So, we have shown analytically ($eq^n(\ref{j4})$), that probability is conserved in the system, in the incipient limit of black hole formation. 

\section{Extremal case}
For the extremal case, $|Q|=M$. From $eq^n(\ref{R-ti})$ for $R_\pm=|Q|=M$ (the event horizon), we obtain the classical behaviour of the shell as,
\begin{align}
R_f = R_\pm + \frac{1}{\left[\frac{t_f}{R_\pm^2}\pm\frac{1}{R_0-R_\pm}\right]}. \label{Rf-ti2}
\end{align}
Like $eq^n(\ref{Rf-ti1})$, $eq^n(\ref{Rf-ti2})$ also suggests that classically, the collapsing shell is infinitely red-shifted for an asymptotic observer. \\

For the extremal case,  for $R\rightarrow R_H$,
\begin{align}
&R(t) = R_\pm + \frac{1}{\left[\frac{t}{R_\pm^2}\pm\frac{1}{R_0-R_\pm}\right]}, \nonumber\\
leading \ to, \ & D \sim \frac{1}{\left(\frac{t}{R_\pm}\pm\frac{R_\pm}{R_0-R_\pm}\right)^2}. \label{D-i4}
\end{align}
Following previous arguments, here one has,
\begin{align}
D&= 
\begin{cases}
1, \ \ \ \ \ \ \ \ \ \ \ \ \ \ \ \ \ \ \ \ \ for \ \ \ t\in(-\infty,0) & \\
\frac{1}{\left(\frac{t}{R_\pm}\pm\frac{R_\pm}{R_0-R_\pm}\right)^2}, \ \ \ \ for \ \ \ t\in(0,t_f) & \\
\frac{1}{\left(\frac{t_f}{R_\pm}\pm\frac{R_\pm}{R_0-R_\pm}\right)^2}, \ \ \ for \ \ \ t\in(t_f,\infty). 
\label{D-choice2}
\end{cases} 
\end{align}  
The corresponding $\omega(\eta(t))$, $\Omega(t)$ and $\Omega(t_f)$ are
\begin{align}
\omega(\eta(t)) = &\left(\frac{t}{R_\pm}\pm\frac{R_\pm}{R_0-R_\pm}\right)\omega_0, \label{omega4} \\
\Omega(t) = &\left(\left.\frac{\partial\eta}{\partial t}\right|_{t>0}\right)\omega(\eta) = \frac{1}{\left(\frac{t}{R_\pm}\pm\frac{R_\pm}{R_0-R_\pm}\right)}\omega_0, \label{omega5} \\
\Omega(t_f) = & \ \frac{1}{\left(\frac{t_f}{R_\pm}\pm\frac{R_\pm}{R_0-R_\pm}\right)}\omega_0. \label{omega6}
\end{align}
The rest of the analysis is similar to the non-extremal case with the above frequencies replacing the previous ones in the corresponding expressions.

\section{Conclusion}
So it has now been shown analytically and comprehensively that the black hole radiation is processed with a unitary evolution. This is accomplished using both facets of unitarity, namely the density matrix consideration as well as the conservation of probability consideration. \\

The Schr\"{o}dinger-like wave equations that we wrote look similar to a minisuperspace version of Wheeler-DeWitt equations\cite{dewitt}. Interestingly, such equations have a present resurgence, in the context of issues regarding unitarity\cite{sridip1, sridip2, sridip3}. However, how the Wheeler-deWitt formalism brings leads to the preservation of unitariy in the models is an issue left to ponder. \\

Saini and Stojkovic\cite{saini1} showed that black hole radiation is a unitary process, for a Schwarzchild black hole, from the density matrix consideration through numerical estimates. We worked with a more general metric, the Reissner-Nordstr\"{o}m metric, and the Schwarzchild results can be trivially recovered from this work. \\

The calculations on unitarity are all in the incipient limit, the limit of formation of the back hole. So it does not really take care of the complete black hole evaporation process. However, if unitarity is preserved in this limit, it should be valid at every instant of time. In fact, there is a claim by Wallace\cite{wallace1} that the information loss paradox is not related to the black hole evaporation, but rather to the formation and its existence, as addressed by the incipient limit (see also \cite{page1, mathur1, polchinski1}). 
\\

In saying this, we further emphasize that, what we have shown in this paper is that black hole radiation is unitary even in a non-globally hyperbolic spacetime. This has significant implications to the resolution of the information loss paradox. \\

It should also be emphasized that this is the first time where unitarity of black hole radiation is checked for an RN spacetime, which is globally non-hyperbolic to start with\cite{lambert}, that is, even as a static spacetime. 

\section*{Acknowledgements}
AD would like to thank the Department of Science and Technology, Government of India for providing the INSPIRE-SHE scholarship which helped immensely in this research work.

\section*{Appendix}
\subsection*{\bf Alternate motivation for $S_{shell}$}
Here we present a different action than $S_{shell}$. We will call it $S_{new}$. We will further show that in the incipient limit it will give rise to $\mathcal{H}_{shell}$ and $\Pi_{shell}$. Since we know that the shell behaves like a relativistic particle, we define the new action to be,
\begin{align}
S_{new} = &-\int d\tau \ M = -\int dT \ \frac{M}{T_\tau}, \nonumber\\
= &-4\pi\sigma\int dT \ R^2\left[1-2\pi\sigma R\sqrt{1-R_T^2}\right] \nonumber\\ 
&-\int dT \ \frac{Q^2}{2R}\sqrt{1-R_T^2}, \nonumber\\
= &-4\pi\sigma\int dt \ R^2\left[\sqrt{D-\frac{1-D}{D}R_t^2}-2\pi\sigma R\sqrt{D-\frac{R_t^2}{D}}\right] \nonumber\\ 
&-\int dt \ \frac{Q^2}{2R}\sqrt{D-\frac{R_t^2}{D}}\label{s1}.
\end{align} 
Then,
\begin{align}
\mathcal{L}_{new} = &-4\pi\sigma R^2\left[\sqrt{D-\frac{1-D}{D}R_t^2}-2\pi\sigma R\sqrt{D-\frac{R_t^2}{D}}\right] \nonumber\\
&-\frac{Q^2}{2R}\sqrt{D-\frac{R_t^2}{D}} \label{lnew}, \\
\Pi_{new} = & \ \frac{\partial\mathcal{L}_{new}}{\partial R_t} \nonumber\\
= & \ \frac{4\pi\sigma R^2R_t}{\sqrt{D}}\left[\frac{1-D}{\sqrt{D^2-(1-D)R_t^2}}-\frac{2\pi\sigma R}{\sqrt{D^2-R_t^2}}\right] \nonumber\\
& + \frac{Q^2}{2R}\frac{R_t}{\sqrt{D}\sqrt{D^2-R_t^2}} \label{pnew},\\
\mathcal{H}_{new} = & \ \Pi_{new}R_t - \mathcal{L}_{new} \nonumber\\
= & \ 4\pi\sigma D^{3/2}R^2\left[\frac{1}{\sqrt{D^2-(1-D)R_t^2}} - \frac{2\pi\sigma R}{\sqrt{D^2-R_t^2}}\right] \nonumber\\
&+\frac{Q^2}{2R}\frac{D^{3/2}}{\sqrt{D^2-R_t^2}} \label{hnew}. 
\end{align}
Now in the incipient limit we have,
\begin{align}
&\mathcal{H}_{new} = \frac{4\pi D^{3/2}\mu R^2}{\sqrt{D^2-R_t^2}} \label{hi}, \\
&\Pi_{new} = \frac{4\pi\mu R^2 R_t}{\sqrt{D}\sqrt{D^2-R_t^2}} \label{pi}, 
\end{align}
where, $\mu:=\sigma\left(1-2\pi\sigma R_H+\frac{Q^2}{8\pi\sigma R_H^3}\right)$. Observe that these are the exact same equations we had obtained before in this incipient limit.

\subsection*{\bf Derivation of the Schr\"{o}dinger-like wave equation from the Wheeler-deWitt equation}
The Wheeler-deWitt equation for a closed universe to which our system belongs to is given by,
\begin{align}
H_{tot}\Psi_{tot} &= 0 \label{wd1},
\end{align}
where $H_{tot}$ is the total Hamiltonian given as,
\begin{align}
H_{tot} &= H_{sys}+H_{obs} \label{H_t},
\end{align}
and $\Psi_{tot}[X^\mu,g_{\mu\nu},\Phi,\mathcal{O}]$ is the total wavefunctional with all the constituents of the system including the observer's degrees of freedom denoted by $\mathcal{O}$. Note that the wavefunctional $\Psi_{tot}$ is a functional only of the fields.\\

We make an assumption that any weak interactions between the observer and the shell-metric-scalar system is contained in $H_{sys}$. Now this, along with the assumption that evolution of the shell-metric-scalar system and that the observer are independent of each other, we can argue that the total wavefunctional is separable and can be written as a sum over eigenstates as,
\begin{align}
\Psi_{tot} &= \sum_k c_k \Psi^k_{sys}(sys,t_{obs}) \Psi^k_{obs}(\mathcal{O},t_{obs}) \label{psi_t},
\end{align}
where $k$ labels the eigenstates $c_k$'s are complex coefficients and $t_{obs}$ is the observer's time coordinate.\\
\\
Based on the above assumptions, we can argue that the observer will have his/her own evolution independent of the dynamics of the system and hence, his/her wavefunction $\Psi^k_{obs}(\mathcal{O},t_{obs})$ is assumed to satisfy the usual Schr\"{o}dinger-like wave equation given as,
\begin{align}
H_{obs}\Psi^k_{obs} &= i\frac{\partial\Psi^k_{obs}}{\partial t_{obs}} \label{obs-wave}.
\end{align}
Now let us go back to $eq^n(\ref{wd1})$, which implies,
\begin{align*}
&\sum_k c_k(H_{sys}+H_{obs})\Psi^k_{sys}\Psi^k_{obs} = 0, \\
leading \ to, \ & \sum_k c_k(H_{sys}\Psi^k_{sys}\Psi^k_{obs}+H_{obs}\Psi^k_{obs}\Psi^k_{sys}) = 0, \\
leading \ to, \ & \sum_k c_k(H_{sys}\Psi^k_{sys}\Psi^k_{obs}) = -\sum_k c_k\left(i\frac{\partial\Psi^k_{obs}}{\partial t_{obs}}\Psi^k_{sys}\right) \\ 
&(by \ eq^n(\ref{obs-wave})).
\end{align*} 
Now integrating the above equation w.r.t. $t_{obs}$ gives,
\begin{align*}
\sum_k c_k &\left[\int^{\infty}_0 dt_{obs} \ (H_{sys}\Psi^k_{sys}\Psi^k_{obs})\right] \\ 
= &-\sum_k c_k\left[\int^{\infty}_0 dt_{obs}\left(i\frac{\partial\Psi^k_{obs}}{\partial t_{obs}}\Psi^k_{sys}\right)\right], \\
= &-\sum_k c_k\left[\left.(\Psi^k_{sys}\Psi^k_{obs})\right|_0^{\infty}-\int_0^{\infty}dt_{obs}\left(i\frac{\partial\Psi^k_{sys}}{\partial t_{obs}}\Psi^k_{obs}\right)\right], \\
= &-\left.\Psi_{tot}\right|_0^{\infty} + \sum_k c_k\left[\int_0^{\infty}dt_{obs}\left(i\frac{\partial\Psi^k_{sys}}{\partial t_{obs}}\Psi^k_{obs}\right)\right] \\ 
&(by \ eq^n(\ref{psi_t})), \\
= &\sum_k c_k\left[\int_0^{\infty}dt_{obs}\left(i\frac{\partial\Psi^k_{sys}}{\partial t_{obs}}\Psi^k_{obs}\right)\right] \\ 
&(as \ \Psi \ is \ t_{obs}-independent).
\end{align*}
Above equation implies,
\begin{align}
\sum_k c_k\left[\int_0^{\infty} dt_{obs}\left(H_{sys}\Psi_{sys}^k-i\frac{\partial\Psi^k_{sys}}{\partial t_{obs}}\right)\Psi_{obs}^k\right] = 0 \label{ckpsi}.
\end{align}
Then for arbitrary states $\Psi_k^{sys}$ and since $c_k$'s are independent of each other, $eq^n(\ref{ckpsi})$ can only hold if the integrand is zero but since $\Psi_{obs}^k$ is not identically zero for all $k$. This implies,
\begin{align}
H_{sys}\Psi^k_{sys} &= i\frac{\partial\Psi^k_{sys}}{\partial t_{obs}} \label{Schro1}.
\end{align}
Thus, the shell-metric-scalar system's wavefunctional $\Psi_{sys}^k$ also satisfies its own Schr\"{o}dinger-like wave equation. We can neglect the subscript $sys$ and the superscript $sys$ and would write $eq^n(\ref{Schro1})$ as,
\begin{align}
H\Psi &= i\frac{\partial\Psi}{\partial t_{obs}} \label{Schro2},
\end{align}
where $H$ is the Hamiltonian and $\Psi$ is the wavefunctional of the shell-metric-scalar system.\\

Following usual minisuperspace arguments, we can truncate the field degrees of freedom to a finite subset and could consider the minisuperspace version of the Wheeler-deWitt equation. This truncation is useful and in the process, we do not lose any useful inputs of the system as long as we keep all the field degrees of freedom important to the analysis. So, since the shell exhibits spherical symmetry by assumption, all fields are assumed to respect spherical symmetry which is a reasonable assumption. So, the shell is described only by the radial degree of freedom denoted by $R(t_{obs})$.\\

Since we are working in the semi-classical regime, which means we are only interested in the quantum effects resulting from the quantization of the scalar field $\Phi$ in the presence of a classical background metric of the shell (where the shell is treated classically), without loss of generality, we can say that, $\mathcal{H}_{\Phi}$ (Hamiltonian for the massless scalar field) is the total Hamiltonian of the system, i.e., $\mathcal{H}_{\Phi}\equiv\mathcal{H}_{sys}$.\\
\\
So for an asymptotic observer, $eq^n(\ref{Schro2})$ reduces to,
\begin{align}
\mathcal{H}_{\Phi}\Psi_{\Phi} = i\frac{\partial\Psi_{\Phi}}{\partial t} \label{Schro3}.
\end{align}
Note that in $eq^n(\ref{Schro3})$, $\Psi_\Phi$ is a wavefunctional which we will be solving for, but this will be equivalent to solving a time-dependent Schr\"{o}dinger equation for a wavefunction, $\Psi(\lbrace a_k\rbrace,t)\equiv\Psi_\Phi$, which is dependent on a set of infinite variables $\lbrace a_k\rbrace$ (the modes) and $t$. Furthermore, it belongs to an infinite dimensional Hilbert space.

\subsection*{\bf Computation of $c_n$}
Let us compute the $c_n$'s explicitly. We know that,
\begin{align}
\psi(b,t) = \sum_n c_n(t)\phi_n(b), \label{psigen}
\end{align}
So, from the overlap integral we have,
\begin{align}
c_n = &\int db \ \phi_n^{*}\psi = \left(\frac{\alpha^2\Omega_f}{\pi^2\zeta^2}\right)^{1/4}\frac{e^{i\chi(\eta)}}{\sqrt{2^n n!}} \nonumber\\
&\int db \ \exp\left[-\frac{\alpha\Omega_f b^2}{2}+i\left(\frac{\zeta_\eta}{\zeta}+\frac{i}{\zeta^2}\right)\frac{\alpha b^2}{2}\right]H_n\left(\sqrt{\alpha\Omega_f}b\right) \label{cn1}, \\
= &\left(\frac{1}{\Omega_f\pi^2\zeta^2}\right)^{1/4}\frac{e^{i\chi(\eta)}}{\sqrt{2^n n!}} \nonumber\\
&\int dx \ \exp\left[-\frac{x^2}{2}+\frac{x^2}{2}\frac{i}{\Omega_f}\left(\frac{\zeta_\eta}{\zeta}+\frac{i}{\zeta^2}\right)\right]H_n(x) \nonumber\\ 
&(with, \ x:=\sqrt{\alpha\Omega_f}b),  \nonumber\\
= &\left(\frac{1}{\Omega_f\pi^2\zeta^2}\right)^{1/4}\frac{e^{i\chi(\eta)}}{\sqrt{2^n n!}}\int dx \ e^{-Px^2/2} H_n(x) \nonumber\\
&\left(with, \ P:=1-\frac{i}{\Omega_f}\left(\frac{\zeta_\eta}{\zeta}+\frac{i}{\zeta^2}\right)\right) \label{cn2} \\
= &\left(\frac{1}{\Omega_f\pi^2\zeta^2}\right)^{1/4}\frac{e^{i\chi(\eta)}}{\sqrt{2^n n!}} I_n \nonumber\\ 
&\left(with, \ I_n:=\int dx \ e^{-Px^2/2} H_n(x)\right) \label{cn3}.
\end{align}
To compute $I_n$, let us consider the following generating function for the $H_n(x)$,
\begin{align}
&J(z) = \int dx \ e^{-Px^2/2}e^{-z^2+2zx} = \sqrt{\frac{2\pi}{P}}e^{-z^2(1-2/P)}, \nonumber\\
since, \ &e^{-z^2+2zx} = \sum_{n=0}^\infty\frac{z^n}{n!}H_n(x), \nonumber\\
&\int dx \ e^{-Px^2/2}H_n(x) = \left.\frac{d^n}{dz^n}J(z)\right|_{z=0}, \nonumber\\
thus, \ &I_n = \sqrt{\frac{2\pi}{P}}\left(1-\frac{2}{P}\right)^{n/2}H_n(0), \nonumber\\
as, \ &H_n(0) = \begin{cases}
(-1)^{n/2}\sqrt{2^n n!}\frac{(n-1)!!}{\sqrt{n!}}, \ \ \ \ for \ \ \ even \ n & \\
0, \ \ \ \ \ \ \ \ \ \ \ \ \ \ \ \ \ \ \ \ \ \ \ \ \ \ \ \ \ for \ \ \ odd \ n. &
\end{cases}\nonumber
\end{align}
Thus we have,
\begin{align}
&c_n = \begin{cases} 
\frac{(-1)^{n/2}e^{i\chi}}{(\Omega_f\zeta^2)^{1/4}}\sqrt{\frac{2}{P}}\left(1-\frac{2}{P}\right)^{n/2}\frac{(n-1)!!}{\sqrt{n!}}, \ \ \ \ for \ \ \ even \ n & \\   
0, \ \ \ \ \ \ \ \ \ \ \ \ \ \ \ \ \ \ \ \ \ \ \ \ \ \ \ \ \ \ \ \ \ \ \ \ \ \ \ \ \ \ \ \ for \ \ \ odd \ n. &
\end{cases} \label{cn4} 
\end{align}

\subsection*{\bf Explicit computation of $Tr(\hat{\rho}_f)$}
We know that,
\begin{align}
Tr(\hat{\rho}_f) = \frac{2}{\sqrt{\Omega_f\zeta^2}|P|}\frac{1}{\sqrt{1-\left|1-\frac{2}{P}\right|^2}} \label{tr1}.
\end{align}
To compute $P$ explicitly, let us give the solution of, 
\begin{align}
\zeta_{\eta\eta} + \omega^2(\eta)\zeta = \frac{1}{\zeta^3}, \label{dif1}
\end{align}
as,
\begin{align}
&\zeta = \frac{1}{\sqrt{\omega_0}}\sqrt{\epsilon^2+\varepsilon^2} \label{z1}, \\
&\zeta_\eta = \frac{1}{\omega_0\zeta}(\epsilon\epsilon_\eta+\varepsilon\varepsilon_\eta) \label{z2}, 
\end{align}
where in terms of Bessel's functions, we have,
\begin{align}
&\epsilon = \frac{\pi u_0}{2}[Y_0(2\omega_0)J_1(u_0)-J_0(2\omega_0)Y_1(u_0)] \label{d1}, \\
&\varepsilon = \frac{\pi u_0}{2}[Y_1(2\omega_0)J_1(u_0)-J_1(2\omega_0)Y_1(u_0)] \label{d2}, \\
&\epsilon_\eta = -\pi\omega_0^2[Y_0(2\omega_0)J_0(u_0)-J_0(2\omega_0)Y_0(u_0)] \label{d3}, \\
&\varepsilon_\eta = -\pi\omega_0^2[Y_1(2\omega_0)J_0(u_0)-J_1(2\omega_0)Y_0(u_0)] \label{d4},
\end{align}
where, $u_0:=2\omega_0\sqrt{1-\eta}$.\\
\\
Now, substituting the definition of $P$ ($eq^n(\ref{cn2})$) in $eq^n(\ref{tr1})$, we have (using {\itshape{Mathematica}}),
\begin{align}
Tr(\hat{\rho}_f) = \frac{|\zeta^2\Omega_f|}{\sqrt{\zeta^2\Omega_f}\sqrt{-\Im[\zeta^2\Omega_f]\Re[\zeta\zeta_\eta]+(1+\Im[\zeta\zeta_\eta])\Re[\zeta^2\Omega_f]}} \label{trmath}.
\end{align}
Now, as $\Omega_f$, $\zeta$ and $\zeta_\eta$ are real (as is evident from $eq^ns(\ref{z1}-\ref{d4})$) , we get from $eq^n(\ref{trmath})$,
\begin{align}
Tr(\hat{\rho}_f) = 1 \label{trf}.
\end{align}


\begin{thebibliography}{100}
\bibitem{hawking1}
S.W. Hawking, Commun. Math. Phys. {\bf 43}, 199 (1975).

\bibitem{hawking2} 
S. W. Hawking,
Phys. Rev. D {\bf 14}, 2460 (1976).

\bibitem{wallace1} 
D. Wallace,
\textcolor{blue}{gr-qc/1710.03783v2} (2017).

\bibitem{mathur1}
S. D. Mathur,
Class. Quant. Grav. {\bf 26}, 224001 (2009).

\bibitem{polchinski1} 
J. Polchinski,
\textcolor{blue}{hep-th/1609.04036v1} (2016).

\bibitem{unruh1}
W.G. Unruh and R.M. Wald,
Phys. Rev. D {\bf 52}, 2176 (1995).

\bibitem{unruh2}
W.G. Unruh and R.M. Wald, 
Rept. Prog. Phys. {\bf 80}, 092002 (2017).

\bibitem{stojkovic1} 
T. Vachaspati, D. Stojkovic and L.M. Krauss,
Phys. Rev. D {\bf 76}, 024005 (2007).

\bibitem{vachaspati1} 
T. Vachaspati,
Class. Quant. Grav. {\bf 26}, 215007 (2009).

\bibitem{saini1} 
A. Saini and D. Stojkovic,
Phys. Rev. Lett. {\bf 114}, 111301 (2015).

\bibitem{saini2}
A. Saini and D. Stojkovic,
Phys. Rev. D {\bf 97}, 025020 (2018).

\bibitem{birkhoff1} 
G. Birkhoff and R. Langer,
{\it Relativity and Modern Physics},
Harvard University Press, Boston, 1923.

\bibitem{heusler1} 
M. Heusler,
\textit{Black Hole Uniqueness Theorems}, 
Cambridge University Press, 1996.

\bibitem{israel1} 
W. Israel,
Nuovo Cimento {\bf 44B}, 1 (1966).

\bibitem{lopez1} 
C. A. L\'{o}pez,
Phys. Rev. D {\bf 38}, 3662 (1988).

\bibitem{wang1} 
J.E. Wang, E. Greenwood and D. Stojkovic, 
Phys. Rev. D {\bf 80}, 124027 (2009).

\bibitem{greenwood1} 
E. Greenwood,
JCAP 1001 002 (2010).

\bibitem{dantas1} 
C. M. A. Dantas, I. A. Pedrosa and B. Baseia,  
Phys. Rev. A {\bf 45}, 1320 (1992).

\bibitem{lewis1} 
H. R. Lewis,
J. Math. Phys. {\bf 9}, 1976 (1968).

\bibitem{lewis2} 
H. R. Lewis. and W. B. Riesenfeld, 
J. Math. Phys. {\bf 10}, 1458 (1969).

\bibitem{pedrosa1} 
I. A. Pedrosa,  
J. Math. Phys. {\bf 28}, 2662 (1987).

\bibitem{kolopanis1} 
M. Kolopanis and T. Vachaspati,
Phys. Rev. D {\bf 87} 085041 (2013).

\bibitem{sakurai1} 
J.J. Sakurai and J. Napolitano, 
\textit{Modern Quantum Mechanics, 2nd Edition}, 
Pearson Education Limited, 2014.

\bibitem{dewitt} 
B.S. DeWitt, Phys. Rev. {\bf 160}, 1113 (1967).

\bibitem{sridip1}  
S. Pal and N. Banerjee, 
Phys. Rev. D {\bf 90}, 104001 (2014).

\bibitem{sridip2} 
S. Pal and N. Banerjee, Phys. Rev. D {\bf 91} 044042 (2015).

\bibitem{sridip3}
S. Pal and N. Banerjee,
J. Math. Phys. {\bf 57}, 122502 (2016).

\bibitem{page1} 
D. N. Page,
Phys. Rev. Lett. {\bf 71}, 3743-3746 (1993).

\bibitem{lambert}
P-H. Lambert,
PoS Modave2013 {\bf 001} (2013).

\end{thebibliography}
\end{document}